\documentclass[journal]{IEEEtran}
\usepackage{epsfig,graphics,epsf,color,amsmath,balance,cite,amsfonts,multirow,stfloats,algorithm,algorithmic,mathrsfs,amssymb,subfigure}
\usepackage[unicode,colorlinks, linkcolor=black,anchorcolor=black,citecolor=black]{hyperref}
\makeatletter
\renewcommand{\maketag@@@}[1]{\hbox{\m@th\normalsize\normalfont#1}}
\makeatother
\newtheorem{proposition}{Proposition}

\begin{document}

\title{Robust ISAC Transceiver Beamforming Design under Low-Resolution AD/DA Converters}

\author{Tiantian Xu,~Zhenyao He,~\IEEEmembership{Student Member,~IEEE,}~Jindan Xu,~\IEEEmembership{Member,~IEEE,}~Wei Xu,~\IEEEmembership{Fellow,~IEEE,}\\~Jianfeng Wang, and Derrick Wing Kwan Ng,~\IEEEmembership{Fellow,~IEEE}

\thanks{

Tiantian Xu, Zhenyao He and Wei Xu are with the National Mobile Communications Research Laboratory, Southeast University, Nanjing 210096, China (e-mail: xutiantian@seu.edu.cn; hezhenyao@seu.edu.cn; wxu@seu.edu.cn). (\textit{Corresponding author: Wei Xu.})

Jindan Xu is with the School of Electrical and Electronics Engineering, Nanyang Technological University, Singapore 639798 (e-mail: jindan.xu@ntu.edu.sg).

Jianfeng Wang is with the Lenovo Research, Lenovo Group Ltd., Beijing 100094, China (e-mail: wangjf20@lenovo.com).

Derrick Wing Kwan Ng is with the School of Electrical Engineering and Telecommunications, University of New South Wales, Sydney, NSW 2052, Australia (e-mail: w.k.ng@unsw.edu.au).
}
}

\maketitle

\begin{abstract}
In this letter, we investigate the robust beamforming design for an integrated sensing and communication (ISAC) system featuring low-resolution digital-to-analog converters (DACs) and analog-to-digital converters (ADCs). Taking into account quantization noise, we aim at maximizing the radar signal-to-quantization-plus-noise ratio (SQNR) while guaranteeing the minimum required signal-to-quantization-plus-interference-plus-noise ratio (SQINR) for communication users. To address this nonconvex design problem, we first examine a scenario involving a point target and uniform-resolution DACs, where the globally optimal solution is obtained by applying the semidefinite relaxation (SDR) technique. For more general scenarios, including those with mixed-DACs and/or an extended target, we develop a low-complexity majorization-minimization (MM)-based algorithm to tackle the problem iteratively. Compared to the non-robust algorithm, the proposed algorithm demonstrates improved detection performance under practical quantization. Simulation results confirm the robustness and efficacy of our proposed algorithm in low-resolution quantization scenarios. 
\end{abstract}


\section{Introduction}
The forthcoming sixth-generation (6G) wireless systems are expected to support both high-rate communication and high-precision sensing. The dramatic escalation in both data traffic and the growing number of connected devices has consequently exacerbated concerns regarding system costs and energy consumption \cite{WXu2023,FLiu2022}. In this context, the integrated sensing and communication (ISAC) paradigm has attracted considerable interest, which seamlessly combines communication and sensing into a unified system, thereby improving the efficiency of spectrum and power \cite{ZTE}.

For ISAC systems, efficient transmission design is crucial for achieving the desired communication and sensing performance. As concerns associated with energy consumption in wireless networks intensify, energy saving has emerged as a principle objective in ISAC transmission design. There exist several studies focused on beamforming design to reduce the transmit power consumption \cite{ZHe2023,FLiu2018}.
However, these works do not tackle the issue of high circuit power consumption.
As carrier frequencies rise and the number of antennas grows, wireless systems face significant challenges in terms of substantial hardware cost and circuit power consumption. 
Notably, the power consumption of digital-to-analog converters (DACs) and analog-to-digital converters (ADCs) increases exponentially with the quantization resolution, posing a critical bottleneck for large-scale system deployment. A pragmatic solution is to employ low-resolution DACs/ADCs. However, this approach introduces non-negligible nonlinear distortion to the signals.

Considerable research has been conducted on the communication and radar systems adopting low-resolution quantizers. For instance, in \cite{JXu2019}, the impact of low-resolution ADCs on the achievable rate in a communication network was analyzed. Besides, utilizing the additive quantization noise model (AQNM), \cite{JChoi2022} and \cite{ZCheng2021} focused on the beamforming design for low-resolution quantized communication and radar systems, respectively. Recently, several studies have also explored the low-resolution quantization in ISAC systems. Specifically, \cite{ZCheng ISAC,XYu2022,WYKeung2025} addressed the transmit sequence design for ISAC systems employing one-bit DACs. Furthermore, the transceiver optimization for ISAC systems with one-bit DACs and ADCs was investigated in \cite{QLin2025} and \cite{BWang2024}.
Despite these fruitful advancements, existing research on ISAC systems has primarily focused on the one-bit quantization, leaving the application of few-bit DACs/ADCs in ISAC systems relatively unexplored.

In light of the above discussions, we investigate the robust transceiver beamforming design for an ISAC system incorporating low-resolution DACs and ADCs in this letter. We focus on the few-bit quantization scenarios where the number of quantization bits exceeds one. Our objective is to maximize the radar signal-to-quantization-plus-noise ratio (SQNR) while adhering to constraints on the communication signal-to-quantization-plus-interference-plus-noise ratio (SQINR) and power consumption. To address the nonconvex design problem, we begin by examining a simplified point target scenario with uniform-resolution DACs at the base station (BS).
In this case, we reformulate the problem and apply the semidefinite relaxation (SDR) to obtain a globally optimal solution. For more general scenarios involving mixed-DACs and/or an extended target, we further propose a majorization-minimization (MM)-based algorithm to acquire a locally optimal solution. Simulation results demonstrate the effectiveness of the proposed robust ISAC beamforming schemes.

\textit{Notations}: Vectors and matrices are denoted by bold lowercase and bold uppercase letters, respectively. $(\cdot)^T$, $(\cdot)^H$, $\mathrm {Tr}(\cdot)$ and $\mathrm{rank}(\cdot)$ stand for the transpose, conjugate transpose, trace and rank of a matrix, respectively.
$\mathbb{C}$ denotes the set of complex numbers. $\mathbb{E}\{\cdot\}$ is the expectation operation. 
$ \mathcal{R}\{\cdot\}$ extracts the real part of the input. $\left | \cdot  \right |$ returns the absolute value of a scalar. $ \left \| \cdot \right \| $ denotes the Euclidean norm. $\mathcal{CN}(\boldsymbol{\mu} ,\mathbf{R} )$ represents the circularly symmetric complex Gaussian distribution with mean $\boldsymbol{\mu}$ and covariance matrix $\mathbf{R}$. $\mathrm{diag}(\mathbf{A})$ forms a diagonal matrix from the diagonal entries of $\mathbf{A}$, while $\mathrm{diag}(a_1,\dots a_n)$ creates a diagonal matrix with specified entries.

\section{System Model and Problem Formulation}

We consider an ISAC BS equipped with $N_T$ transmit antennas and $N_R$ receive antennas, which simultaneously serves $K$ downlink communication users while performing monostatic radar sensing. The transmit and receive arrays are co-located and both arranged as uniform linear arrays (ULAs) with half-wavelength spacing. To reduce the hardware cost and power consumption, the BS utilizes low-resolution DACs at the transmitter, along with both the BS and users employ low-resolution ADCs at the receivers.

Let $\mathbf{s}_{c} = [s_{c,1}, \ldots, s_{c,K}]^T \in \mathbb{C}^K$ represent the communication symbols sent by the BS to $K$ users and $\mathbf{s}_r \in \mathbb{C}^{N_T}$ denotes the dedicated radar symbols. Then, the digital transmit signal at the BS is given by \cite{XLiu2020}
\small{
\begin{equation}
    \label{eq_x}
    \mathbf{x}=\mathbf{W}_{c} \mathbf{s}_{c}+\mathbf{W}_{r} \mathbf{s}_{r},
\end{equation}
}\normalsize
where $\mathbf{W}_c \!\!\in\! \mathbb{C}^{N_T \times K}$ and $\mathbf{W}_r \!\!\in\! \mathbb{C}^{N_T \times N_T}$ denote the communication and radar beamforming matrices, respectively. Without loss of generality, we assume that the communication and radar symbols are independent and identically distributed with zero mean and unit power, i.e., $ \mathbb{E} \left \{ \mathbf{s}_c \mathbf{s}_c^H \right \} \!\!=\!\! \mathbf{I}_K $, $\mathbb{E} \left \{ \mathbf{s}_r \mathbf{s}_r^H \right \} \!\!=\!\! \mathbf{I}_{N_T}$, and $\mathbb{E}\left \{ \mathbf{s}_c \mathbf{s}_r^H \right \} \!\!=\!\! \mathbf{0}_{K \times N_T}$. Based on these assumptions, the covariance matrix of $\mathbf{x}$ is $\mathbf{R}_\mathbf{x}=\mathbf{W}_{r}\mathbf{W}_{r}^H\!+\!\mathbf{W}_{c} \mathbf{W}_{c}^H$ \cite{XLiu2020}.
Adopting the AQNM\footnote{AQNM and Bussgang decomposition \cite{Bussgang} are widely utilized to derive linear approximations of quantized signals in research on low-resolution quantization. In comparison, AQNM offers a simpler formulation, but ignores the correlations among the entries of the quantization noise vector. Bussgang decomposition, which accounts for these correlations, provides higher accuracy. However, when the number of quantization bits exceeds one, the Bussgang-based model lacks a closed-form expression \cite{SJacobsson2019}. Hence, we adopt AQNM in this work.} \cite{AKFletcher2007}, we obtain the analog transmit signal quantized through low-resolution DACs, written as
\small{
\begin{equation}
    \label{eq_xq}
    \mathbf{x}_q = \mathbf{A}_t \mathbf{x} + \mathbf{q}_t,
\end{equation}
}\normalsize
where $ \mathbf{A}_t=\mathrm{diag}(\alpha_{t,1},\dots ,\alpha_{t,N_T})\in \mathbb{C}^{N_T \times N_T}$ represents the quantization gain matrix and $ \mathbf{q}_t \in \mathbb{C}^{N_T}$ is the quantization noise, uncorrelated to the input signal to DACs, which follows the distribution $\mathcal{CN}(\mathbf{0},\mathbf{R}_{\mathbf{q}_t}) $. The covariance matrix $\mathbf{R}_{\mathbf{q}_t} \in \mathbb{C}^{N_T \times N_T}$ is expressed as $\mathbf{R}_{\mathbf{q}_{t}}=\mathbf{A}_{t}(\mathbf{I}_{N_T}-\mathbf{A}_{t})\mathrm{diag}(\mathbf{R}_{\mathbf{x}})$.
Besides, $ \alpha_{t,n}=1-\beta_{t,n}, n \in \left \{1,\dots,N_T  \right \}  , $ is the quantization gain of the DAC on the $n$-th radio-frequency (RF) chain, where $ \beta_{t,n}$ is a quantization distortion factor uniquely determined by the number of quantization bits $b_{t,n}$. For $ b_{t,n} \leq 5 $, the values of $\beta_{t,n}$ are shown in \cite[Table I]{ZCheng2021}. For $ b_{t,n} > 5 $, $ \beta_{t,n} $ is approximately expressed as $ \beta_{t,n} = \frac{\sqrt{3}\pi}{2} \cdot 2^{-2b_{t,n}} $ \cite{JChoi2022}. Furthermore, the total transmit power is calculated as $ \mathrm{Tr}\left(\mathbf{A}_{t}\mathbf{R}_{\mathbf{x}}\mathbf{A}_{t}^{H}+\mathbf{A}_{t}(\mathbf{I}_{N_T}-\mathbf{A}_{t})\mathrm{diag}(\mathbf{R}_{\mathbf{x}})\right)$.

The downlink received signal at user $k$ is given by
\small{
\begin{equation}
    \label{eq_yk}
    y_{k}=\mathbf{h}_{k}^{H}\mathbf{x}_{q}+z_{k}=\mathbf{h}_{k}^{H}\mathbf{A}_{t}\mathbf{x}+\mathbf{h}_{k}^{H}\mathbf{q}_{t}+z_{k}, \quad \forall k,
\end{equation}
}\normalsize
where $\mathbf{h}_k\in\mathbb{C}^{N_T}$ is the channel vector between the BS and user $k$, and $z_{k}$ denotes the additive white Gaussian noise with variance $\sigma_{k}^{2}$ at user $k$. We assume perfect channel state information (CSI) at the BS \cite{ZHe2023,JChoi2022,XLiu2020}. The received signal $y_k$ is then quantized at the ADC of user $k$, whose output is given by \cite{AKFletcher2007}
\small{
\begin{equation}
    \label{eq_ykq}
    y_{k,q}=\alpha_{k}y_{k}+q_{k}\!=\!\alpha_{k}\mathbf{h}_{k}^{H}\mathbf{A}_{t}\mathbf{x}+\alpha_{k}\mathbf{h}_{k}^{H}\mathbf{q}_{t}+\alpha_{k}z_{k}+q_{k},\forall k,
\end{equation}
}\normalsize
where $\alpha_{k}=1-\beta_{k}$ is the quantization gain and $\beta_{k}$ denotes the quantization distortion factor, defined similarly as the DAC counterparts. $ q_{k} \in \mathbb{C}$ denotes the quantization noise, following the distribution $\mathcal{CN}(0,r_{q_{k}})$, where $r_{q_{k}}=\alpha_{k}(1-\alpha_{k})r_{k}$ and $r_{k} = \mathbb{E} \{ |y_k|^2 \} =\mathbf{h}_{k}^{H}\mathbf{A}_{t}\mathbf{R}_{\mathbf{x}}\mathbf{A}_{t}^{H}\mathbf{h}_{k}+\mathbf{h}_{k}^{H}\mathbf{R}_{\mathbf{q}_{t}}\mathbf{h}_{k}+\sigma_{k}^{2}$.
Denoting the $k$-th column of $\mathbf{W}_c$ by $\mathbf{w}_k$, we then derive the SQINR of user $k$ by (\ref{gamma_k}) at the top of the next page \cite{JChoi2022}.

\begin{figure*}[hbt]
    \centering
    \vspace*{-15pt}
    {\small
    \begin{align}
        \label{gamma_k}
        \gamma_{k} \!=\!\frac{\alpha_k^2\mathbf{h}_k^H\mathbf{A}_t\mathbf{w}_k\mathbf{w}_k^H\mathbf{A}_t^H\mathbf{h}_k}{\alpha_k^2\mathbf{h}_k^H\mathbf{A}_t(\sum_{i\neq k}\!\mathbf{w}_i\mathbf{w}_i^H\!+\!\mathbf{W}_r\!\mathbf{W}_r^H)\mathbf{A}_t^H\mathbf{h}_k\!+\!\alpha_k^2\mathbf{h}_k^H\mathbf{R}_{\mathbf{q}_t}\mathbf{h}_k\!+\!\alpha_k^2\sigma_k^2\!+\!r_{q_k}} 
        \!=\!\frac{\alpha_{k}\mathbf{h}_{k}^{H}\mathbf{A}_{t}\mathbf{w}_{k}\mathbf{w}_{k}^{H}\mathbf{A}_{t}^{H}\mathbf{h}_{k}}{\mathbf{h}_{k}^{H}\mathbf{A}_{t}(\mathbf{R}_{\mathbf{x}}\!-\!\alpha_{k}\mathbf{w}_{k}\mathbf{w}_{k}^{H})\mathbf{A}_{t}^{H}\mathbf{h}_{k}\!+\!\mathbf{h}_{k}^{H}\mathbf{R}_{\mathbf{q}_{t}}\mathbf{h}_{k}\!+\!\sigma_{k}^{2}},\forall k.
    \end{align}
    }\normalsize
    \hrulefill
    \vspace*{-13pt}
\end{figure*}

On the other hand, let $ \mathbf{G} \in \mathbb{C}^{N_R \times N_T} $ be the target response matrix (TRM) for sensing \cite{FLiu CRB}. The analog signal at the BS receiver is given by
\small{
\begin{equation}
    \label{eq_yBS}
    \mathbf{y}_{\mathrm{BS}}=\mathbf{G}\mathbf{x}_{q}+\mathbf{z}_{\mathrm{BS}}=\mathbf{G}\mathbf{A}_{t}\mathbf{x}+\mathbf{G}\mathbf{q}_{t}+\mathbf{z}_{\mathrm{BS}},
\end{equation}
}\normalsize
where $ \mathbf{z}_{\mathrm{BS}} \in \mathbb{C}^{N_R} $ denotes the additive white Gaussian noise with covariance matrix $\sigma_r^2 \mathbf{I}_{N_R}$. The digital received signal after ADC quantization is expressed as
\small{
\begin{equation}
    \label{eq_yBSq}
    \mathbf{y}_{\mathrm{BS},q}=\mathbf{A}_r\mathbf{y}_{\mathrm{BS}}\!+\!\mathbf{q}_r=\mathbf{A}_r\mathbf{G}\mathbf{A}_t\mathbf{x}\!+\!\mathbf{A}_r\mathbf{G}\mathbf{q}_t\!+\!\mathbf{A}_r\mathbf{z}_{\mathrm{BS}}\!+\!\mathbf{q}_r.
\end{equation}
}\normalsize
Here, $\mathbf{A}_{r}=\mathrm{diag}(\alpha_{r,1},\dots ,\alpha_{r,N_{R}})\in\mathbb{C}^{N_{R}\times N_{R}}$ is the quantization gain matrix, where $\alpha_{r,n}=1-\beta_{r,n}$ and $\beta_{r,n}$ is the quantization distortion factor of the low-resolution ADC on the $n$-th RF chain of the receiver, following the same relationship with the number of quantization bits as $\beta_{t,n}$. The quantization noise $\mathbf{q}_r\in\mathbb{C}^{N_R}$ satisfies $\mathbf{q}_{r}{\sim}\mathcal{CN}(\mathbf{0},\mathbf{R}_{\mathbf{q}_{r}})$ with $\mathbf{R}_{\mathbf{q}_r}=\mathbf{A}_r(\mathbf{I}_{N_R}-\mathbf{A}_r)\mathrm{diag}(\mathbf{R}_{\mathbf{y}_{\mathrm{BS}}})$ 
and $\mathbf{R}_{\mathbf{y}_{\mathrm{BS}}}=\mathbb{E}\left\{\mathbf{y}_{\mathrm{BS}}\mathbf{y}_{\mathrm{BS}}^H \right\}=\mathbf{G}\mathbf{A}_{t}\mathbf{R}_{\mathbf{x}}\mathbf{A}_{t}^{H}\mathbf{G}^{H}+\mathbf{G}\mathbf{R}_{\mathbf{q}_{t}}\mathbf{G}^{H}+\sigma_{r}^{2}\mathbf{I}_{N_R}$.

We adopt radar SQNR as the performance metric for sensing, owing to its close relationship with detection probability. Similar metrics have been used in \cite{ZCheng2021,QLin2025,MDeng2022}. By denoting the BS receive beamforming vector as $ \mathbf{u} \in \mathbb{C}^{N_R \times 1} $, we can express the radar SQNR at the BS as
\small{
\begin{equation}
    \label{gamma_r}
    \gamma_{r}=\frac{\mathbf{u}^{H}\mathbf{A}_{r}\mathbf{G}\mathbf{A}_{t}\mathbf{R}_{\mathbf{x}}\mathbf{A}_{t}^{H}\mathbf{G}^{H}\mathbf{A}_{r}^{H}\mathbf{u}}{\mathbf{u}^{H}\big(\mathbf{A}_{r}\mathbf{G}\mathbf{R}_{\mathbf{q}_{t}}\mathbf{G}^{H}\mathbf{A}_{r}^{H}+\sigma_{r}^{2}\mathbf{A}_{r}\mathbf{A}_{r}^{H}+\mathbf{R}_{\mathbf{q}_{r}}\big)\mathbf{u}}.
\end{equation}
}\normalsize
Define $\mathbf{Q}=\mathbf{A}_{r}\mathbf{G}\mathbf{R}_{\mathbf{q}_{t}}\mathbf{G}^{H}\mathbf{A}_{r}^{H}+\sigma_{r}^{2}\mathbf{A}_{r}\mathbf{A}_{r}^{H}+\mathbf{R}_{\mathbf{q}_{r}}$. According to the properties of the generalized Rayleigh quotient \cite{ZHe2023}, $\gamma_r $ achieves its maximum value at $\mathbf{u}=\mathbf{Q}^{-1}\mathbf{A}_{r}\mathbf{GA}_{t}\mathbf{x}$, thereby rendering the radar SQNR $ \gamma_r $ as:
\small{
\begin{equation}
    \label{gamma_r_max}
    \gamma_r^{\max}=\mathrm{Tr}(\mathbf{A}_r\mathbf{G}\mathbf{A}_t\mathbf{R}_\mathbf{x}\mathbf{A}_t^H\mathbf{G}^H\mathbf{A}_r^H\mathbf{Q}^{-1}).
\end{equation}
}\normalsize

Our goal is to design the transmit beamformer for low-resolution quantization scenarios, aiming to enhance sensing performance while ensuring communication performance. Accordingly, we maximize the radar SQNR described in (\ref{gamma_r_max}) under the constraints of transmit power and communication SQINR. The optimization problem is formulated as
\small{
\begin{subequations}
    \label{p1}
    \begin{align}
        \operatorname*{maximize}_{\mathbf{W}_{r},\mathbf{W}_{c}}&~~~\mathrm{Tr}(\mathbf{A}_{r}\mathbf{G}\mathbf{A}_{t}\mathbf{R}_{\mathbf{x}}\mathbf{A}_{t}^{H}\mathbf{G}^{H}\mathbf{A}_{r}^{H}\mathbf{Q}^{-1})& \label{17a} \\
        \mathrm{subject~to}&~~~\gamma_{k}\geq\Gamma_{k},\quad \forall k, & \label{17b} \\
        &~~~\mathrm{Tr}\big(\mathbf{A}_{t}\mathbf{R}_{\mathbf{x}}\mathbf{A}_{t}^{H}\!\!+\!\mathbf{A}_{t}(\mathbf{I}_{N_T}\!\!-\!\mathbf{A}_{t})\mathrm{diag}(\mathbf{R}_{\mathbf{x}})\big)\!\leq\! P, \label{17c}
    \end{align}
\end{subequations}
}\normalsize
where $\Gamma_k$ is the minimum required SQINR threshold for user $k$ and $P$ is the total transmit power budget.

\section{Proposed Solution}

Problem (\ref{p1}) is difficult to solve directly due to its nonconvex objective function and communication SQINR constraints.
Given its complexity, we initially tackle the problem within a basic point target scenario, where the BS employs DACs of uniform resolution.
In this setting, we derive the globally optimal solution to (\ref{p1}). Subsequently, we extend our analysis to more general cases, considering mixed-DACs \cite{ZCheng2021} configurations and extended target scenarios.

\subsection{Beamforming Optimization for Point Target with Uniform-resolution DACs}
The difficulty in solving problem (\ref{p1}) primarily stems from the matrix inversion $ \mathbf{Q}^{-1} $ associated with the beamforming matrices in the objective function. Fortunately, this inversion can be effectively circumvented in the point target scenario with uniform-resolution DACs at the BS.

Since the BS transmitter adopts a uniform quantization resolution, the quantization gain of the DAC on each RF chain is identical, denoted as $ \alpha_t $. Thus, the matrix $ \mathbf{A}_t $ can be simplified to $ \mathbf{A}_t = \alpha_t \mathbf{I}_{N_T} $. In the point target scenario, the sensing channel matrix is written as $ \mathbf{G} = \eta \mathbf{b}(\theta) \mathbf{a}^H(\theta) $, where $ \eta \in \mathbb{C} $ is the reflection coefficient, $ \theta $ is the direction of the point target, and $ \mathbf{a}(\theta) \!\in\! \mathbb{C}^{N_T} $ and $ \mathbf{b}(\theta) \!\in\! \mathbb{C}^{N_R} $ are the transmit and receive steering vectors, respectively, represented as $\mathbf{a}(\theta)\!=\!\left[1,e^{-j\pi\sin\theta},\dots,e^{-j(N_{T}-1)\pi\sin\theta}\right]^{T}$ and $\mathbf{b}(\theta)\!=\!\left[1,e^{j\pi\sin\theta},\dots,e^{j(N_{R}-1)\pi\sin\theta}\right]^{T}$. For brevity, $\mathbf{a}$ and $\mathbf{b}$ are adopted to denote $\mathbf{a} (\theta) $ and $ \mathbf{b} (\theta) $ in the sequel, respectively. Based on these conditions, we can derive an equivalent problem to (\ref{p1}), as shown in the following proposition.

\begin{proposition}\label{prop1}
    In the point target scenario with uniform-resolution DACs at the BS transmitter, problem (\ref{p1}) can be equivalently reformulated as:
    \small{
    \begin{align}
        \label{p2}
        \operatorname*{maximize}_{\mathbf{W}_{r},\mathbf{W}_{c}} \qquad &\mathbf{a}^{H}\mathbf{R}_{\mathbf{x}}\mathbf{a}& \notag \\[-2pt]
        \mathrm{subject~to} \qquad &\gamma_{k}\geq\Gamma_{k}, \quad \forall k, & \notag \\
        &\alpha_{t}\mathrm{Tr}(\mathbf{R_{x}})\leq P.&
    \end{align}
    }\normalsize
\end{proposition}

\begin{IEEEproof}
    Please refer to Appendix \ref{proof:1}.
\end{IEEEproof}

Compared to problem (\ref{p1}), the matrix inversion has been removed from the objective function in problem (\ref{p2}), allowing it to be solved exploiting the SDR technique. Specifically, define $ \mathbf{R}_k = \mathbf{w}_k \mathbf{w}_k^H, \forall k, $ with $\mathrm{rank}(\mathbf{R}_{k})=1,\mathbf{R}_{k}\succeq \mathbf{0},\forall k$, and $\mathbf{R}_{\mathbf{x}}-\sum_{k=1}^{K}\mathbf{R}_{k}\succeq \mathbf{0}$. Then, removing the rank-one constraints from problem (\ref{p2}) yields
\small{
\begin{align}
    \label{p3}
    \operatorname*{maximize}_{\mathbf{R}_{\mathbf{x}}\succeq \mathbf{0}, \{\mathbf{R}_{k}\succeq \mathbf{0}\}_{k=1}^{K}} &&& \mathbf{a}^{H}\mathbf{R_{x}a} & \notag \\[-7pt] 
    \mathrm{subject~to~~~} &&& \left(1+\frac{1}{\Gamma_k}\right)\alpha_k\alpha_t^2\mathbf{h}_k^H\mathbf{R}_k\mathbf{h}_k\ge  \alpha_t^2\mathbf{h}_k^H\mathbf{R}_\mathbf{x}\mathbf{h}_k & \notag \\
    &&& +\alpha_t(1-\alpha_t)\mathbf{h}_k^H\mathrm{diag}(\mathbf{R}_\mathbf{x})\mathbf{h}_k+\sigma_k^2,\quad \forall k, & \notag \\
    &&&\mathbf{R_x}-\sum_{k=1}^K\mathbf{R_k}\succeq\mathbf{0}, & \notag \\
    &&&\alpha_{t}\mathrm{Tr}(\mathbf{R}_{\mathbf{x}})\leq P, &
\end{align}
}\normalsize
which is a semidefinite programming (SDP) that can be optimally solved via numerical convex programming solvers. 
According to \cite[Theorem 1]{XLiu2020}, it holds that for any globally optimal solution $ \mathbf{\widehat{R}}_{\mathbf{x}}, \{\mathbf{\widehat{R}}_{k}\}_{k=1}^{K} $ of problem (\ref{p3}), one can always construct another optimal solution $ \mathbf{\widetilde{R}}_{\mathbf{x}}, \{\mathbf{\widetilde{R}}_{k}\}_{k=1}^{K} $ satisfying the rank-one constraints, i.e., $\mathrm{rank}(\mathbf{R}_{k})\!=\!1,\forall k$, through the transformations $ \mathbf{\widetilde{R}}_{\mathbf{x}}\!=\!\mathbf{\widehat{R}}_{\mathbf{x}},\mathbf{\widetilde{R}}_{k}=( \mathbf{h}^H_k \mathbf{\widehat{R}}_{k} \mathbf{h}_k )^{-1} \mathbf{\widehat{R}}_{k} \mathbf{h}_k \mathbf{h}^H_k \mathbf{\widehat{R}}_{k}^H, \forall k $.
Therefore, we can obtain the globally optimal solution of problem (\ref{p2}) by solving problem (\ref{p3}).

\textit{Remark 1:} When disregarding the quantization effects of the DACs and ADCs at the BS, the radar SNR can be rederived as $ \frac{1}{\sigma_r^2} \mathrm{Tr}(\mathbf{G}\mathbf{R}_\mathbf{x}\mathbf{G}^H) $ following the approach in Section~II. Notably, in the point target scenario, maximizing $ \frac{1}{\sigma_r^2} \mathrm{Tr}(\mathbf{G}\mathbf{R}_\mathbf{x}\mathbf{G}^H) $ is equivalent to maximizing $ \mathbf{a}^{H}\mathbf{R}_{\mathbf{x}}\mathbf{a} $, which aligns exactly with the objective function of problem (\ref{p2}). Therefore, problem (\ref{p2}) shares the same objective function with the radar SNR maximization problem that excludes the BS quantization process.

\subsection{Beamforming Optimization for General Scenarios}
For more general scenarios, such as those involving mixed-DACs, multiple point targets, or an extended target, the forms of $ \mathbf{A}_t $ and $ \mathbf{G}$ become increasingly complex, rendering it more challenging to simplify problem (\ref{p1}) as effectively as before. Therefore, we employ an MM-based algorithm to address problem (\ref{p1}) by solving a series of surrogate problems.

To this end, we first perform a variable substitution and address the nonconvex communication SQINR constraints (\ref{17b}). By defining $\mathbf{V}=[\mathbf{W}_{c},\mathbf{W}_{r}]$, we have $\mathbf{R}_{\mathbf{x}}=\mathbf{V}\mathbf{V}^{H}$ and $\mathbf{w}_{k}=\mathbf{V} \mathbf{e}_{k}$, where $ \mathbf{e}_k $ is a $ K $-dimensional column vector with the $ k $-th entry being 1 and the rest being 0. Then, the objective function (\ref{17a}) becomes
\small{
\begin{equation}
    \label{f_V}
    f(\mathbf{V})=\mathrm{Tr}(\mathbf{V}^{H}\mathbf{A}_{t}^{H}\mathbf{G}^{H}\mathbf{A}_{r}^{H}\mathbf{Q}^{-1}\mathbf{A}_{r}\mathbf{G}\mathbf{A}_{t}\mathbf{V}),
\end{equation}
}\normalsize
where $ \mathbf{Q} $ is related to $ \mathbf{R}_{\mathbf{q}_t} $ and $ \mathbf{R}_{\mathbf{q}_r} $, with $\mathbf{R}_{\mathbf{q}_t}=\mathbf{A}_t(\mathbf{I}_{N_T}-\mathbf{A}_t)\mathrm{diag}(\mathbf{V}\mathbf{V}^H)$ and $\mathbf{R}_{\mathbf{q}_{r}}=\mathbf{A}_{r}(\mathbf{I}_{N_R}-\mathbf{A}_{r})\mathrm{diag}(\mathbf{G}\mathbf{A}_{t}\mathbf{V}\mathbf{V}^{H}\mathbf{A}_{t}^{H}\mathbf{G}^{H}+\mathbf{G}\mathbf{R}_{\mathbf{q}_{t}}\mathbf{G}^{H}+\sigma_{r}^{2}\mathbf{I}_{N_R})$. The SQINR of user $k$ is rewritten as
\small{
\begin{align}
    \label{gamma_k2}
    \gamma_{k}\!= \!\frac{\alpha_{k}\mathbf{h}_{k}^{H}\mathbf{A}_{t}\mathbf{Ve}_{k}\mathbf{e}_{k}^{H}\mathbf{V}^{H}\mathbf{A}_{t}^{H}\mathbf{h}_{k}}{\mathbf{h}_{k}^{H}\!\mathbf{A}_{t}(\mathbf{V}\mathbf{V}^{H}\!\!-\!\alpha_{k}\!\mathbf{V}\mathbf{e}_{k}\mathbf{e}_{k}^{H}\mathbf{V}^{H})\mathbf{A}_{t}^{H}\mathbf{h}_{k}\!+\!\mathbf{h}_{k}^{H}\mathbf{R}_{\mathbf{q}_{t}}\!\mathbf{h}_{k}\!+\!\sigma_{k}^{2}}.
\end{align}
}\normalsize
According to (\ref{gamma_k2}), we further express the communication SQINR constraints (\ref{17b}) as
\footnotesize{
\begin{align}
    \label{SQINR1}
    &\alpha_k\left(1+\frac{1}{\Gamma_k}\right)\left|\mathbf{h}_k^H\mathbf{A}_t\mathbf{V}\mathbf{e}_k\right|^2 \geq \notag \\ 
    &\left \|\mathbf{V}^H\!\mathbf{A}_t^H\mathbf{h}_k\right \|^2\!\!+\!\!\sum_{i=1}^{N_T}\left|h_{k,i}\right|^2\!\alpha_{t,i}\!\left(1\!-\!\alpha_{t,i}\right) \!\left \|\mathbf{v}_i\right \|^2\!+\!\sigma_k^2,~\forall k,
\end{align}
}\normalsize
where $h_{k,i}$ denotes the $i$-th element of $\mathbf{h}_k $, and $\mathbf{v}_i$ represents the conjugate transpose of the $i$-th row of $\mathbf{V}$. Notice that if $\mathbf{w}_{k}=\mathbf{V}\mathbf{e}_{k}$ is optimal, then for any phase $\phi_k$, $\mathbf{w}_ke^{j\phi_k}$ is also optimal. Thus, we can select $\phi_k$ such that $\mathbf{h}_{k}^{H}\mathbf{A}_{t}\mathbf{w}_{k}$, or equivalently $\mathbf{h}_{k}^{H}\mathbf{A}_{t}\mathbf{V}\mathbf{e}_{k}$, becomes a real number. Therefore, (\ref{SQINR1}) can be equivalently transformed into
{\footnotesize
\begin{align}
    \label{SQINR2}
    &\sqrt{\alpha_k\left(1+\frac1{\Gamma_k}\right)}\mathbf{h}_k^H\mathbf{A}_t\mathbf{V}\mathbf{e}_k\geq \notag \\ 
    &\sqrt{\left \| \mathbf{V}^H\mathbf{A}_t^H\mathbf{h}_k  \right \| ^2\!+\!\sum_{i=1}^{N_T}\left|h_{k,i}\right|^2\alpha_{t,i}(1\!-\!\alpha_{t,i})\!\left\|\mathbf{v}_i\right \|^2\!+\!\sigma_k^2},~\forall k,
\end{align}
}\normalsize
which is a set of convex second-order cone constraints. Next, we focus on dealing with the nonconvex objective function $ f(\mathbf{V}) $ in (\ref{f_V}). By leveraging the property that $\mathrm{Tr}(\mathbf{X}\mathbf{Z}^{-1}\mathbf{X}^{H})$ is jointly convex in $ \mathbf{Z} \succ \mathbf{0} $ and $ \mathbf{X} $, we obtain its first-order Taylor series expansion:
{\small
\begin{align}
    \label{ieq}
    \mathrm{Tr}(\mathbf{X}\mathbf{Z}^{-1}\!\mathbf{X}^{H}) \!\! \geq \!\! 2\mathcal{R}\{\mathrm{Tr}(\mathbf{Z}_{0}^{-1}\!\mathbf{X}_{0}^{H}\!\mathbf{X})\} 
     \!\!-\!\! \mathrm{Tr}(\mathbf{Z}_{0}^{-1}\mathbf{X}_{0}^{H}\mathbf{X}_{0}\mathbf{Z}_{0}^{-1}\!\mathbf{Z}),
\end{align}
}\normalsize
where the equality holds when $ \mathbf{X} = \mathbf{X}_0 $ and $ \mathbf{Z} = \mathbf{Z}_0 $. By substituting $\mathbf{X}=\mathbf{V}^{H}\mathbf{A}_{t}^{H}\mathbf{G}^{H}\mathbf{A}_{r}^{H}$ and $\mathbf{Z}=\mathbf{A}_{r}\mathbf{G}\mathbf{R}_{\mathbf{q}_{t}}\mathbf{G}^{H}\mathbf{A}_{r}^{H}+\sigma_{r}^{2}\mathbf{A}_{r}\mathbf{A}_{r}^{H}+\mathbf{R}_{\mathbf{q}_{r}}$, a surrogate function, $ g(\mathbf{V}, \mathbf{V}_m) $, of $f(\mathbf{V})$ can be constructed as follows:
{\small
\begin{align}
    \label{g_VVm}
    f(\mathbf{V}) & \geq g(\mathbf{V},\mathbf{V}_m) \notag \\ 
    &=2\mathcal{R}\{\mathrm{Tr}(\mathbf{Z}_m^{-1}\mathbf{X}_m^H\mathbf{X})\}\!-\!\mathrm{Tr}(\mathbf{Z}_m^{-1}\mathbf{X}_m^H\mathbf{X}_m\mathbf{Z}_m^{-1}\mathbf{Z}),
\end{align}
}\normalsize
where $ \mathbf{V}_m $, $ \mathbf{X}_m $, and $ \mathbf{Z}_m $ denote the values of $ \mathbf{V} $, $ \mathbf{X} $, and $ \mathbf{Z} $ at the $m$-th iteration of the MM algorithm, respectively. It can be observed that $ g(\mathbf{V}, \mathbf{V}_m) $ is a concave function in $ \mathbf{V} $. Thus, we address problem (\ref{p1}) by iteratively solving the following convex problem:
{\small
\begin{align}
    \label{p4}
    \operatorname*{maximize}_\mathbf{V} ~~& g(\mathbf{V},\mathbf{V}_{m}) \notag \\
    \mathrm{subject~to}~~&\mathrm{Tr}\big(\mathbf{A}_{t}\!\mathbf{V}\mathbf{V}^{H}\!\!\mathbf{A}_{t}^{H}\!\!+\!\!\mathbf{A}_{t}(\mathbf{I}_{N_T}\!\!-\!\mathbf{A}_{t})\mathrm{diag}(\mathbf{V}\mathbf{V}^{H})\big)\!\leq\! P, \notag \\
    &(\ref{SQINR2}), 
\end{align}
}\normalsize
which can be tackled by the interior point method with a complexity of $ \mathcal{O}\left ( N_T^6 K^{1.5}+N_T^3 K^{4.5} \right ) $ \cite{KYWang2014}. It can be proven that the proposed MM-based algorithm always converges to a stationary point of problem (\ref{p1}) \cite[Theorem 1]{ZHe MSE}.

\section{Simulation Results}

In this section, we evaluate the performance of the proposed robust beamforming algorithms. Unless specified otherwise, we set $N_T\!=\!16$, $N_R\!=\!16$, $K\!=\!4$, $P\!=\!20~\mathrm{dBm}$, $ \Gamma_k\!=\!5~\mathrm{dB}, \forall k, $ and $\sigma_r^2\!=\!\sigma_k^2\!=\!0~\mathrm{dBm}, \forall k$ \cite{XLiu2020}. 
We assume that the communication channels follow the Rayleigh fading model. For point target scenarios, the target is assumed to be located at an angle of $\theta\!=\!40^\circ$ and the reflection coefficient satisfies $|\eta|^{2}\!\!=\!\!-10~\mathrm{dB}$ \cite{ZHe2023}. For extended target scenarios, following \cite{FLiu CRB}, each entry of $\mathbf{G}$ is assumed to be mutually independent and follows the distribution $\mathcal{CN}(0,\sigma_{g}^{2})$, where $\sigma_{g}^{2}\!\!=\!\!-10~\mathrm{dB}$. 
For performance comparisons, in addition to the proposed method (``Proposed robust''), we also simulate three benchmark schemes: 1) the non-robust design that neglects the quantization effects (``Non-robust''); 2) the CRB-min design in \cite{FLiu CRB} (``CRB-min''); 3) the scheme without communication SQINR constraints, serving as an upper bound for radar performance (``Radar-only'').
The problem formulation of the non-robust algorithm is identical to the proposed algorithm, with the only difference being the assumption of no quantization loss during the system model construction. 

\begin{figure}[tbp]
  \centering
  \subfigure[Point target.]{\includegraphics[width=0.488\columnwidth]{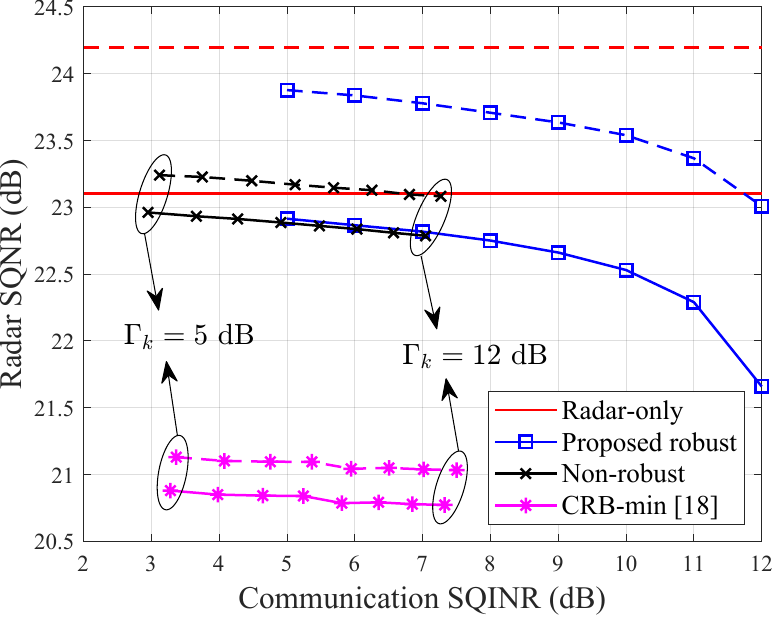}}\hspace{1mm}
  \subfigure[Extended target.]{\includegraphics[width=0.488\columnwidth]{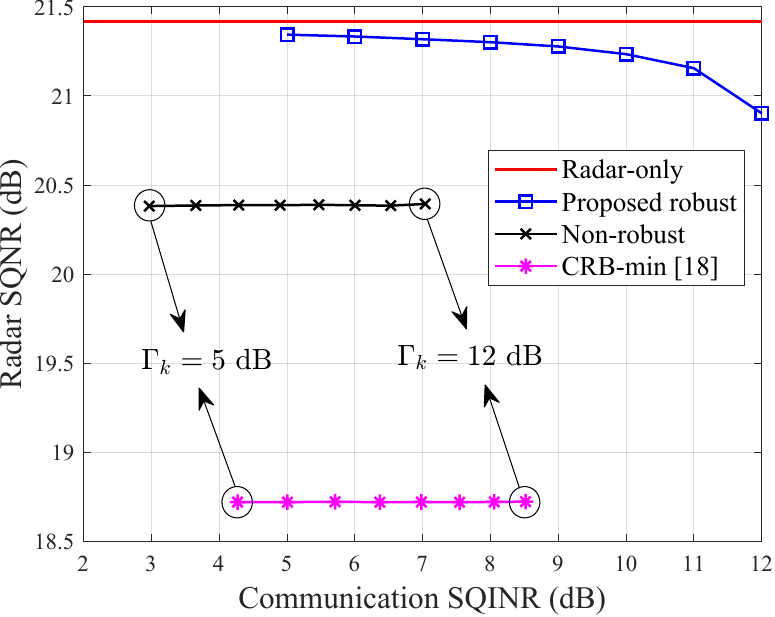}}
  \caption{Radar SQNR versus communication SQINR. Solid lines: the BS employs uniform 3-bit DACs. Dashed lines: the BS employs mixed-DACs (14 pairs of 3-bit and 2 pairs of 10-bit).}
  \label{fig:Gamma}
  \vspace{-3mm}
\end{figure}

We illustrate the relationship between the radar SQNR and the communication SQINR for different schemes in Fig.~\ref{fig:Gamma}, where the communication SQINR threshold is set between $5\text{--}12$ dB. All users employ 3-bit ADCs.
The horizontal axis represents the average communication SQINR of the $K$ users achieved by each algorithm. 
As shown in Fig.~\ref{fig:Gamma}(a), in the point target scenario, the radar SQNR of the three algorithms decreases with the increase of the communication SQINR, reflecting the non-trivial trade-off between communication and sensing performance. 
It can be observed that the proposed robust algorithm outperforms other schemes. However, in the case with uniform-resolution DACs, the radar SQNR obtained from the proposed robust and non-robust algorithm is nearly the same for identical communication SQINR. This is because the simplified problem (\ref{p2}) obtained by the robust algorithm shares the equivalent objective function as the non-robust algorithm in this scenario, as explained in Remark 1.
Furthermore, the simulation curve for the non-robust algorithm exhibits a leftward shift, indicating that it fails to satisfy the set communication SQINR threshold. This is due to its neglect of the low-resolution ADCs at users, resulting in a performance loss. In contrast, our proposed algorithm effectively mitigates this issue.
As depicted in Fig.~\ref{fig:Gamma}(b), the simulation results for the extended target scenario show a similar trend to those in the point target scenario with mixed-DACs. The improvement in radar and communication performance of the proposed algorithm can also be observed.

\begin{figure}[tbp]
  \centering
  \subfigure[Point target with mixed-DACs.]{\includegraphics[width=0.488\columnwidth]{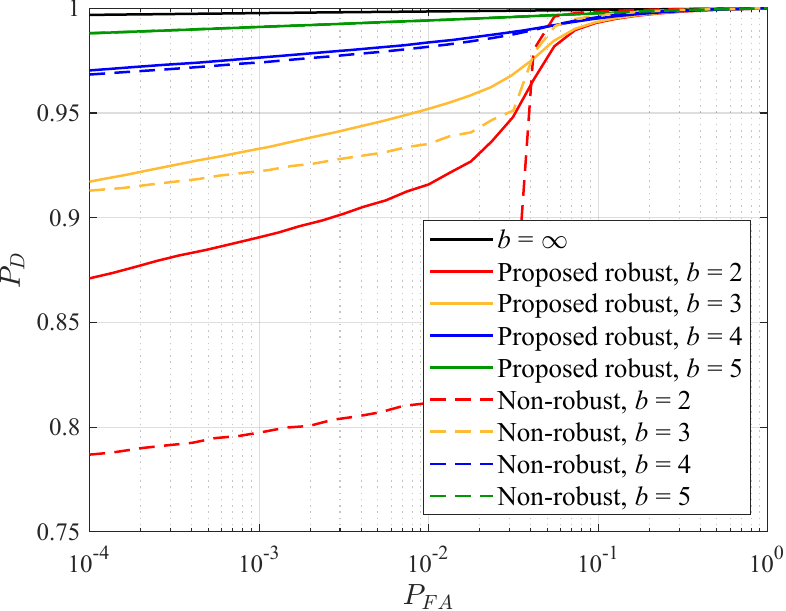}}\hspace{1mm}
  \subfigure[Extended target.]{\includegraphics[width=0.488\columnwidth]{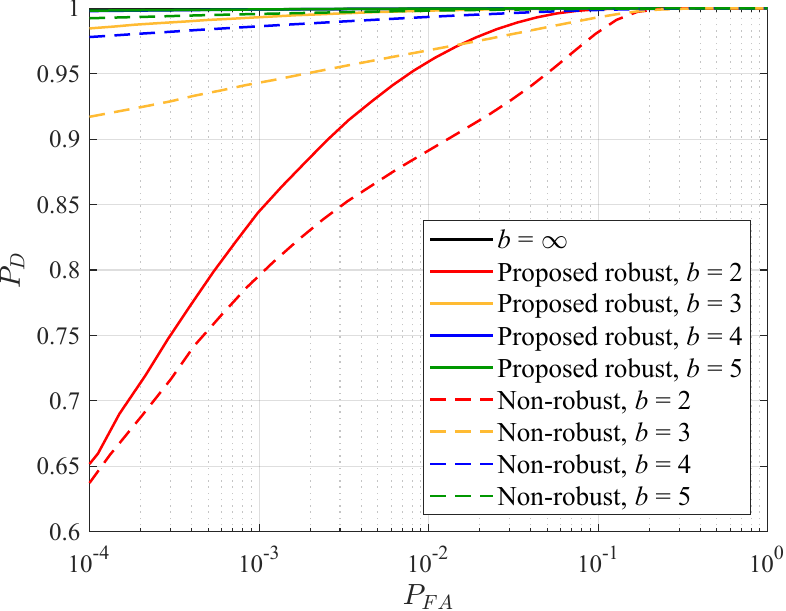}}
  \caption{ROC performance comparison. $P_D$: detection probability. $P_{F\!A}$: false alarm probability.}
  \label{fig:ROC}
  \vspace{-3mm}
\end{figure}

In Fig.~\ref{fig:ROC}, we compare the receiver operating characteristic (ROC) of the proposed robust algorithm with that of the non-robust algorithm under practical quantization. In this simulation, uniform mid-rise quantization and an energy detector are used. To isolate the impact of low-resolution quantization at the BS on detection performance, we assume that the users employ infinite-resolution ADCs. For the point target scenario with mixed-DACs, the BS is equipped with two pairs of 10-bit DACs, while the remaining DACs and ADCs adopt $b$-bit quantization. For the extended target scenario, the BS employs $b$-bit quantization for all DACs and ADCs. As can be observed, the proposed algorithm outperforms the non-robust algorithm in detection performance, demonstrating its effectiveness under practical low-resolution quantization.
Additionally, in Fig.~\ref{fig:ROC}(a), $P_D$ decreases rapidly at high $P_{FA}$ values when $b=2,3$. This is because low-resolution quantization results in a more concentrated distribution of $\mathbf{u}^{H} \mathbf{y}_{\mathrm{BS},q}$ at lower energy levels, thus degrading the detection performance.

\begin{figure}[tbp]
  \centering
  \subfigure[]{\includegraphics[width=0.488\columnwidth]{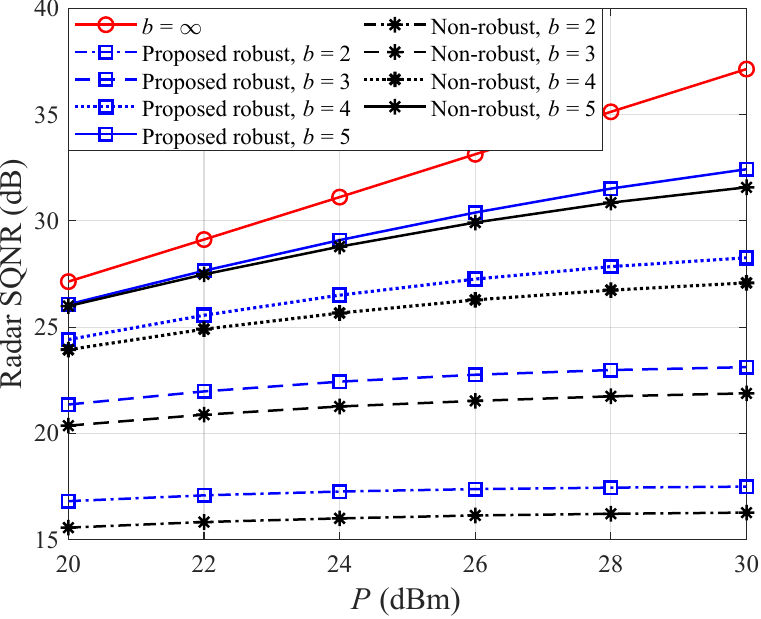}}\hspace{1mm}
  \subfigure[]{\includegraphics[width=0.488\columnwidth]{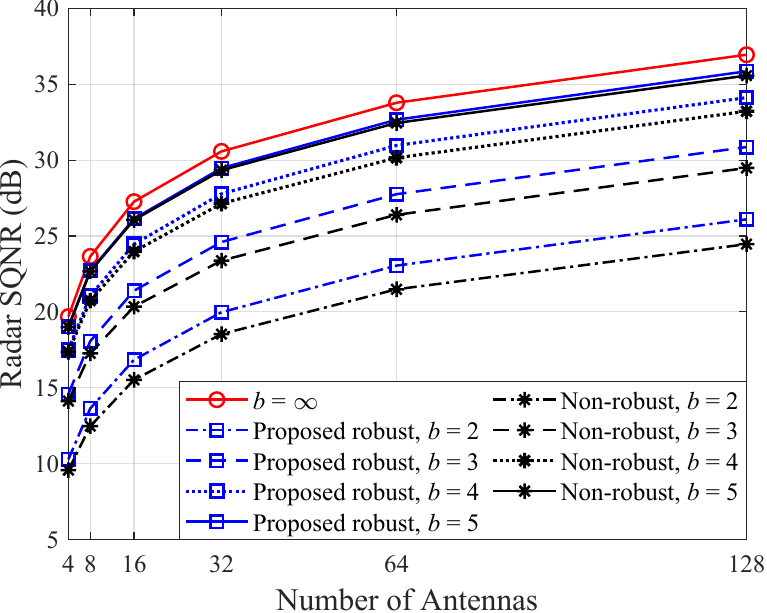}}
  \caption{(a) Radar SQNR versus the transmit power; (b) Radar SQNR versus the number of antennas when $ K = 2 $, with the number of transmit and receive antennas at BS being equal.}
  \label{fig:PN}
  \vspace{-3mm}
\end{figure}

Fig.~\ref{fig:PN} depicts the radar SQNR versus the transmit power and the number of BS antennas in the extended target scenario. The settings for quantization bits remain the same as Fig.~\ref{fig:ROC}(b).
The results indicate that the proposed algorithm shows significant gains for $ b = 2,3,4 $, but there are diminishing returns in the improvement for $ b = 5 $, which is expected due to its proximity to the ideal scenario of infinite-resolution quantization. Additionally, as shown in Fig.~\ref{fig:PN}(a), the gain of the proposed algorithm at $ b = 5 $ improves as $ P $ increases, suggesting that the performance bottleneck is jointly influenced by both the transmit power and the number of quantization bits.
From Fig.~\ref{fig:PN}(b), it can be observed that the gain of the proposed algorithm improves with the number of antennas, but the growth rate diminishes progressively. This is because when the number of antennas is sufficiently large, the performance improvement in mitigating the adverse effects of low-resolution quantization through beamforming approaches reaches saturation.

\begin{figure}[tbp]
  \centering
  \subfigure[]{\includegraphics[width=0.488\columnwidth]{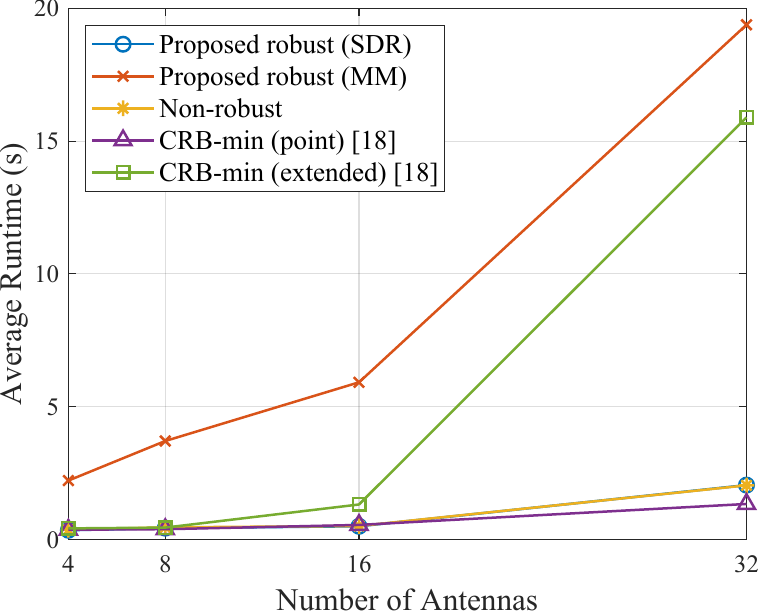}}\hspace{1mm}
  \subfigure[]{\includegraphics[width=0.482\columnwidth]{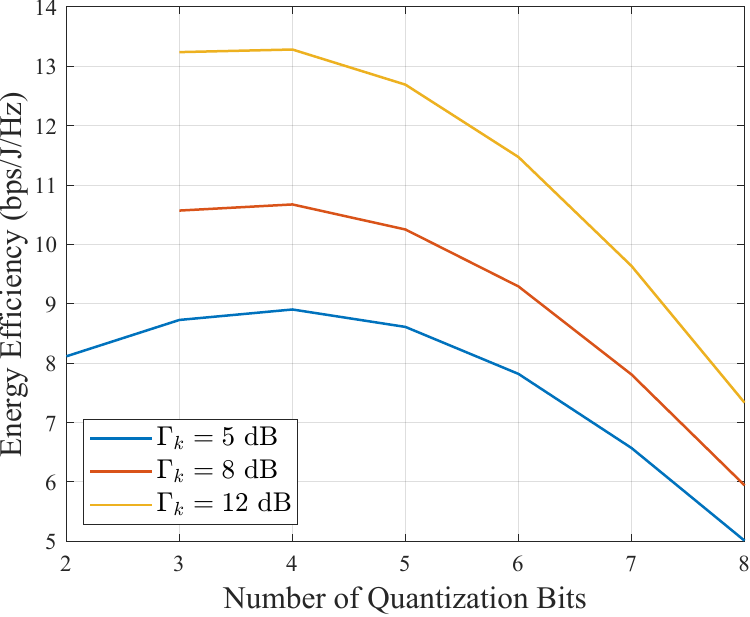}}
  \caption{(a) Average runtime versus the number of antennas. (b) Energy efficiency versus the number of quantization bits.}
  \label{fig:time_EE}
  \vspace{-3mm}
\end{figure}

Fig.~\ref{fig:time_EE}(a) shows the average runtime of the four algorithms, all of which increase with the number of antennas.
In addition, the proposed MM-based algorithm typically converges within 10 iterations, which enhances the robustness of the ISAC system under low-resolution quantization with an acceptable time cost.
Fig.~\ref{fig:time_EE}(b) presents the energy efficiency versus the number of quantization bits in the extended target scenario. All quantizers utilize the same quantization resolution. 
The energy efficiency is defined as 
{\small
\begin{align}
    \label{EE}
    \eta = \frac{{\sum\limits_{k = 1}^{K}{\log_{2}\left( 1 + \gamma_{k} \right)}} + {\log_{2}\left( 1 + \gamma_{r} \right)}}{P_{BS} + KP_{UE}}, 
\end{align}
}\normalsize
where the BS power consumption $ P_{BS} = P_{LO} + N_{T}\left( {P_{RF} + 2P_{DAC}} \right) + N_{R}\left( {P_{RF} + 2P_{ADC}} \right) + \kappa^{- 1}P$, and the user power consumption $ P_{UE} = P_{LO} + P_{RF} + 2P_{ADC}$. Here, $ P_{LO}$, $ P_{RF}$, and $ P_{DAC}$ refer to the power consumption of the local oscillator, RF chain, and DAC, respectively. $\kappa $ denotes the power amplifier efficiency. Their values or calculation methods are referenced in \cite{JChoi2022}. The power consumption of ADC $ P_{ADC}$ is calculated as detailed in \cite{TLiu2019}. 
The results show that maximum energy efficiency is achieved at $b=4$, indicating that low-resolution quantization provides a favorable trade-off between power consumption and performance.

\section{Conclusion}
This letter investigated the robust beamforming optimization for an ISAC system equipped with low-resolution DACs and ADCs, where the radar SQNR was maximized under communication SQINR and power constraints. Under the point target scenario with uniform-resolution DACs, we reformulated the problem equivalently and solved it optimally. 
Subsequently, we proposed an MM-based algorithm to tackle more general scenarios, including those involving mixed-DACs and extended targets. Simulation results demonstrated that the proposed algorithm achieved significantly improved communication and detection performance compared to the non-robust approach.

\begin{appendices}
\section{Proof of Proposition ~\ref{prop1}}
\label{proof:1}
Substituting $ \mathbf{A}_t = \alpha_t \mathbf{I}_{N_T} $ into the power constraint in (\ref{17c}), it reduces to $\alpha_t\mathrm{Tr}(\mathbf{R}_\mathbf{x})\leq P$. We first proceed to prove that when problem (\ref{p1}) reaches its optimal solution, constraint (\ref{17c}) must be active such that $ \mathrm {Tr}(\mathbf{R}_\mathbf{x}) = \frac{P}{\alpha_t}$.

Suppose that $\{\mathbf{W}_{c},\mathbf{W}_{r}\}$ is an optimal solution to problem (\ref{p1}), and that $ \mathrm{Tr}( \mathbf{R}_{\mathbf{x}} ) = \frac{\kappa P}{\alpha_{t}}$, where $
0 < \kappa < 1$. Then, one can construct $\mathbf{W}{}'_{c} = \frac{1}{\sqrt{\kappa}}\mathbf{W}_{c},\mathbf{W}{}'_{r} = \frac{1}{\sqrt{\kappa}}\mathbf{W}_{r}$, which satisfies $ \mathrm{Tr}(\mathbf{R}{}'_\mathbf{x}) = \mathrm{Tr}(\mathbf{W}{}' _r \mathbf{W}{}'^H_r + \mathbf{W}{}' _c \mathbf{W}{}'^H_c) = \frac{P}{\alpha_t}$.

The relationship between the communication SQINR $\gamma{}'_k$ and $\gamma_k$ can be derived as
{\small
\begin{align}
    \gamma{}'_{k}&= \frac{\alpha_{k}\mathbf{h}_{k}^{H}\mathbf{A}_{t}\mathbf{w}{}'_{k}{\mathbf{w}{}'^{H}_{k}}\mathbf{A}_{t}^{H}\mathbf{h}_{k}}{\mathbf{h}_{k}^{H}\mathbf{A}_{t}( {\mathbf{R}{}'_{\mathbf{x}} - \alpha_{k}\mathbf{w}{}'_{k}{\mathbf{w}{}'^{H}_{k}}} )\mathbf{A}_{t}^{H}\mathbf{h}_{k} + \mathbf{h}_{k}^{H}\mathbf{R}{}'_{\mathbf{q}_{t}}\mathbf{h}_{k} + \sigma_{k}^{2}} \notag  \\
    &= \frac{\alpha_{k}\mathbf{h}_{k}^{H}\mathbf{A}_{t}\mathbf{w}_{k}\mathbf{w}_{k}^{H}\mathbf{A}_{t}^{H}\mathbf{h}_{k}}{\mathbf{h}_{k}^{H}\mathbf{A}_{t}( {\mathbf{R}_{\mathbf{x}} - \alpha_{k}\mathbf{w}_{k}\mathbf{w}_{k}^{H}} )\mathbf{A}_{t}^{H}\mathbf{h}_{k} + \mathbf{h}_{k}^{H}\mathbf{R}_{\mathbf{q}_{t}}\mathbf{h}_{k} + \kappa\sigma_{k}^{2}} \notag \\
    &> \gamma_{k} \geq \Gamma_{k}, \quad \forall k.
\end{align}
}\normalsize
Therefore, $\{\mathbf{W}{}'_{c},\mathbf{W}{}'_{r}\} $ satisfies constraint (\ref{17b}) and constitutes a feasible solution to problem (\ref{p1}). 

Denote the objective function of problem (\ref{p1}) by $f_{\mathrm{obj}}( {\mathbf{W}_{c},\mathbf{W}_{r}} )$. Then,
{\small
\begin{align}
    f_{\mathrm{obj}}( {\mathbf{W}{}'_{c},\mathbf{W}{}'_{r}} ) &= \mathrm{Tr}( {\mathbf{A}_{r}\mathbf{G}\mathbf{A}_{t}\mathbf{R}{}'_{\mathbf{x}}\mathbf{A}_{t}^{H}\mathbf{G}^{H}\mathbf{A}_{r}^{H}{\mathbf{Q}{}'}^{-1}} ) \notag \\
    &= \frac{1}{\kappa}\mathrm{Tr}( {\mathbf{A}_{r}\mathbf{G}\mathbf{A}_{t}\mathbf{R}_{\mathbf{x}}\mathbf{A}_{t}^{H}\mathbf{G}^{H}\mathbf{A}_{r}^{H}{\mathbf{Q}{}'}^{- 1}} ),
\end{align}
}\normalsize
where
{\footnotesize
\begin{align}
    \mathbf{Q}{}' &= \mathbf{A}_{r}\mathbf{G}\mathbf{R}{}'_{\mathbf{q}_{t}}\mathbf{G}^{H}\mathbf{A}_{r}^{H} + \sigma_{r}^{2}\mathbf{A}_{r}\mathbf{A}_{r}^{H} \\
    &~~~+ \mathbf{A}_{r}( \mathbf{I}_{N_{R}} \!- \!\mathbf{A}_{r} )\mathrm{diag}( \mathbf{G}\mathbf{A}_{t}\mathbf{R}{}'_{\mathbf{x}}\mathbf{A}_{t}^{H}\mathbf{G}^{H} \!+\! \mathbf{G}\mathbf{R}{}'_{\mathbf{q}_{t}}\mathbf{G}^{H} \!+\! \sigma_{r}^{2}\mathbf{I}_{N_{R}} ) \notag  \\
    &= \frac{1}{\kappa}( {\mathbf{Q} - \sigma_{r}^{2}\mathbf{A}_{r}} ) + \sigma_{r}^{2}\mathbf{A}_{r} \notag \\
    &= \frac{1}{\kappa}\lbrack {\mathbf{Q} + \left( {\kappa - 1} \right)\sigma_{r}^{2}\mathbf{A}_{r}} \rbrack.
\end{align}
}\normalsize
Therefore, $ f_{\mathrm{obj}}( {\mathbf{W}{}'_{c},\mathbf{W}{}'_{r}} )$ can be further expressed as
{\small
\begin{align}
    &~~~~f_{\mathrm{obj}}( {\mathbf{W}{}'_{c},\mathbf{W}{}'_{r}} ) \notag \\
    &= \mathrm{Tr}\left\{ {\mathbf{A}_{r}\mathbf{G}\mathbf{A}_{t}\mathbf{R}_{\mathbf{x}}\mathbf{A}_{t}^{H}\mathbf{G}^{H}\mathbf{A}_{r}^{H}\lbrack {\mathbf{Q} + \left( {\kappa - 1} \right)\sigma_{r}^{2}\mathbf{A}_{r}} \rbrack^{- 1}} \right\}. 
\end{align}
}\normalsize

We proceed to prove that $ f_{\mathrm{obj}}( {\mathbf{W}{}'_{c},\mathbf{W}{}'_{r}} )\ge f_{\mathrm{obj}}( {\mathbf{W}_{c},\mathbf{W}_{r}} )$. Subtracting the two terms leads to
{\small
\begin{align}
    \label{eq12}
    f_{\mathrm{obj}}( {\mathbf{W}{}'_{c},\mathbf{W}{}'_{r}} ) \!-\! f_{\mathrm{obj}}( {\mathbf{W}_{c},\mathbf{W}_{r}} ) 
    \!\triangleq\! \mathrm{Tr}( {\mathbf{A}_{r}\mathbf{G}\mathbf{A}_{t}\mathbf{R}_{\mathbf{x}}\mathbf{A}_{t}^{H}\mathbf{G}^{H}\mathbf{A}_{r}^{H}\mathbf{\Xi}} ).
\end{align}
}\normalsize
Here, 
{\small
\begin{align}
    \mathbf{\Xi} &\triangleq  \lbrack {\mathbf{Q} + \left( {\kappa - 1} \right)\sigma_{r}^{2}\mathbf{A}_{r}} \rbrack^{- 1}-\mathbf{Q}^{-1} \notag \\
    &\overset{(a)}{=} - \mathbf{Q}^{- 1}\left( {\mathbf{Q}^{- 1} + \frac{1}{\left( {\kappa - 1} \right)\sigma_{r}^{2}}\mathbf{A}_{r}^{- 1}} \right)^{- 1}\mathbf{Q}^{- 1},
\end{align}
}\normalsize
where the equality in (a) follows directly from the application of the Woodbury formula.
Since $\mathbf{Q}^{H} = \mathbf{Q} $, it follows that $\mathbf{\Xi}^{H} = \mathbf{\Xi} $. Furthermore, $\mathbf{\Xi}$ is conjugate congruent to $-\left( {\mathbf{Q}^{- 1} + \frac{1}{\left( {\kappa - 1} \right)\sigma_{r}^{2}}\mathbf{A}_{r}^{- 1}} \right)^{- 1}$. According to the Searle set of identities, $-\left( {\mathbf{Q}^{- 1} + \frac{1}{\left( {\kappa - 1} \right)\sigma_{r}^{2}}\mathbf{A}_{r}^{- 1}} \right)^{- 1}$ can be further transformed as
{\small
\begin{align}
    - ( {\mathbf{Q}^{- 1} \!\!+\! \frac{1}{\left( {\kappa \!-\! 1} \right)\sigma_{r}^{2}}\mathbf{A}_{r}^{- 1}} )^{- 1} 
    \!\!\!&= \left( {1 \!-\! \kappa} \right)\sigma_{r}^{2}\mathbf{Q}\left( {\mathbf{Q} \!+\! \left( {\kappa \!-\! 1} \right)\sigma_{r}^{2}\mathbf{A}_{r}} \right)^{- 1}\!\!\mathbf{A}_{r} \notag  \\
    &\triangleq \left( {1 - \kappa} \right)\sigma_{r}^{2}\mathbf{Q}\mathbf{\Pi}^{- 1}\mathbf{A}_{r},
\end{align}
}\normalsize
where $\mathbf{\Pi} \triangleq \mathbf{Q} + \left( {\kappa - 1} \right)\sigma_{r}^{2}\mathbf{A}_{r} = \mathbf{A}_{r}\mathbf{G}\mathbf{R}_{\mathbf{q}_{t}}\mathbf{G}^{H}\mathbf{A}_{r}^{H} + \mathbf{A}_{r}\left( {\mathbf{I}_{N_{R}} - \mathbf{A}_{r}} \right)\mathrm{diag}\left( {\mathbf{G}\mathbf{A}_{t}\mathbf{R}_{\mathbf{x}}\mathbf{A}_{t}^{H}\mathbf{G}^{H} + \mathbf{G}\mathbf{R}_{\mathbf{q}_{t}}\mathbf{G}^{H}} \right) + \kappa\sigma_{r}^{2}\mathbf{A}_{r} $. Since $\mathbf{\Pi}$ is the sum of a positive semidefinite matrix and a positive definite matrix, we have $\mathbf{\Pi} \succ \mathbf{0}$, which implies that $\mathbf{\Pi}^{-1} \succ \mathbf{0}$. Moreover, given that $\left( {1 - \kappa} \right)\sigma_{r}^{2} > 0, \mathbf{Q} \succ \mathbf{0}$ and $ \mathbf{A}_{r} \succ \mathbf{0}$, all eigenvalues of the matrix $\left( {1 - \kappa} \right)\sigma_{r}^{2}\mathbf{Q}\mathbf{\Pi}^{- 1}\mathbf{A}_{r}$ are strictly positive. As a result, $- \left( {\mathbf{Q}^{- 1} + \frac{1}{\left( {\kappa - 1} \right)\sigma_{r}^{2}}\mathbf{A}_{r}^{- 1}} \right)^{- 1} \succ \mathbf{0}$. Since $\mathbf{\Pi}$ is conjugate congruent to $- \left( {\mathbf{Q}^{- 1} + \frac{1}{\left( {\kappa - 1} \right)\sigma_{r}^{2}}\mathbf{A}_{r}^{- 1}} \right)^{- 1}$, it follows that $\mathbf{\Xi} \succ \mathbf{0}$.

Since $\mathbf{A}_{r}\mathbf{G}\mathbf{A}_{t}\mathbf{R}_{\mathbf{x}}\mathbf{A}_{t}^{H}\mathbf{G}^{H}\mathbf{A}_{r}^{H}\succeq \mathbf{0}$ and $\mathbf{\Xi} \succ \mathbf{0}$, we have $\mathrm{Tr}( {\mathbf{A}_{r}\mathbf{G}\mathbf{A}_{t}\mathbf{R}_{\mathbf{x}}\mathbf{A}_{t}^{H}\mathbf{G}^{H}\mathbf{A}_{r}^{H}\mathbf{\Xi}} ) \ge 0$. Substituting this result into (\ref{eq12}), we obtain $ f_{\mathrm{obj}}( {\mathbf{W}{}'_{c},\mathbf{W}{}'_{r}} )\ge f_{\mathrm{obj}}( {\mathbf{W}_{c},\mathbf{W}_{r}} )$, which implies that there exists a feasible solution to problem (\ref{p1}) that is better than $\{\mathbf{W}_{c},\mathbf{W}_{r}\}$, contradicting the initial assumption. Therefore, the optimal solution to problem (\ref{p1}) must satisfy $ \mathrm{Tr}( \mathbf{R}_{\mathbf{x}} ) = \frac{ P}{\alpha_{t}}$. 

We now focus on simplifying the objective function of problem (\ref{p1}). Under the conditions $\mathbf{G}=\eta\mathbf{b}\mathbf{a}^{H}$ and $ \mathbf{A}_t = \alpha_t \mathbf{I}_{N_T} $, the matrix $ \mathbf{Q} $ is further expressed as
{\footnotesize
\begin{align}
    \label{eq_Q}
    \mathbf{Q}\!=\!\mathbf{A}_r \! \left ( \! \left |\eta \right |^2\!\varepsilon\mathbf{b}\mathbf{b}^H\!\!\mathbf{A}_r^H\!\!+\!\!\left |\eta \right |^2\!\left (\zeta\!\left (\mathbf{W}_r,\!\!\mathbf{W}_c\right )\!\!+\!\!\varepsilon\right )\!(\mathbf{I}_{N_R}\!-\!\mathbf{A}_r\!)\!\!+\!\!\sigma_r^2\mathbf{I}_{N_R}\!\!\right )\!,
\end{align}
}\normalsize
where $\zeta(\mathbf{W}_{r},\mathbf{W}_{c})=\alpha_{t}^{2}\mathbf{a}^{H}\mathbf{R}_{\mathbf{x}}\mathbf{a}$ and $\varepsilon=\mathbf{a}^{H}\mathbf{R}_{\mathbf{q}_{t}}\mathbf{a}=\alpha_{t}(1-\alpha_{t})\mathrm{Tr}(\mathbf{R}_{\mathbf{x}})$. Since $ \mathrm {Tr}(\mathbf{R}_\mathbf{x}) = \frac{P}{\alpha_t}$, $ \varepsilon $ can be simplified to $\varepsilon=(1-\alpha_{t})P$, which is a constant independent of $ \mathbf{W}_{r} $ and $ \mathbf{W}_{c} $. Let us define $ \mathbf{B} = \left |\eta \right |^2\left (\zeta\left (\mathbf{W}_r,\mathbf{W}_c\right )+\varepsilon\right )(\mathbf{I}_{N_R}-\mathbf{A}_r)+\sigma_r^2\mathbf{I}_{N_R}$. According to the Sherman-Morrison formula, the inverse matrix of $ \mathbf{Q} $ is computed as
{\footnotesize
\begin{equation}
    \label{invQ}
    \mathbf{Q}^{-1}=\left(\mathbf{B}^{-1}-\frac{|\eta|^2\varepsilon\mathbf{B}^{-1}\mathbf{b}\mathbf{b}^H\mathbf{A}_r^H\mathbf{B}^{-1}}{1+|\eta|^2\varepsilon\mathbf{b}^H\mathbf{A}_r^H\mathbf{B}^{-1}\mathbf{b}}\right)\mathbf{A}_r^{-1}.
\end{equation}
}\normalsize
Since $ \mathbf{B}$ is a diagonal matrix, its inverse can be easily obtained. Given (\ref{invQ}), the objective function (\ref{17a}) is modified to
{\footnotesize
\begin{align}
    \label{obj_fun}
    &\mathrm{Tr}(\mathbf{A}_r\mathbf{G}\mathbf{A}_t\mathbf{R}_\mathbf{x}\mathbf{A}_t^H\mathbf{G}^H\!\mathbf{A}_r^H\mathbf{Q}^{-1})& \notag \\
    =& |\eta|^{2}\!\zeta( {\mathbf{W}_{r},\!\mathbf{W}_{c}} )\mathrm {Tr}( {\mathbf{A}_{r}\mathbf{b}\mathbf{b}^{H}\!\!\mathbf{A}_{r}^{H}\!\mathbf{Q}^{- 1}} \!)
    \!=\!\frac{|\eta|^{2}\zeta(\mathbf{W}_{r},\mathbf{W}_{c})}{\frac{1}{\mathbf{b}^{H}\!\mathbf{A}_{r}^{H}\mathbf{B}^{-1}\mathbf{b}}\!+\!|\eta|^{2}\varepsilon}.
\end{align}
}\normalsize
Given that $ \mathbf{A}_{r} $ and $\mathbf{B}$ are both diagonal matrices, $\mathbf{A}_{r}^{H}\mathbf{B}^{- 1}$ is also diagonal. Substituting $ \mathbf{b} = \left[1,e^{j\pi\sin\theta},\dots,e^{j(N_{R}-1)\pi\sin\theta}\right]^{T}$ into $\mathbf{b}^{H}\mathbf{A}_{r}^{H}\mathbf{B}^{-1}\mathbf{b}$, it can be rewritten as $\sum_{i=1}^{N_R}\frac{\alpha_{r,i}}{|\eta|^2(\zeta(\mathbf{W}_r,\mathbf{W}_c)+\varepsilon)(1-\alpha_{r,i})+\sigma_r^2}$.
Thus, the objective function is further reduced to
{\footnotesize
\begin{align}
    \label{obj_fun2}
    &\mathrm{Tr}(\mathbf{A}_{r}\mathbf{G}\mathbf{A}_{t}\mathbf{R}_{\mathbf{x}}\mathbf{A}_{t}^{H}\mathbf{G}^{H}\mathbf{A}_{r}^{H}\mathbf{Q}^{-1})=|\eta|^{2}\zeta(\mathbf{W}_{r},\mathbf{W}_{c})& \notag \\
    &  \times  \bigg ( \frac{1}{\sum_{i=1}^{N_{R}}\frac{\alpha_{r,i}}{|\eta|^{2}(\zeta(\mathbf{W}_{r},\mathbf{W}_{c})+\varepsilon)\left(1-\alpha_{r,i}\right)\!+\!\sigma_{r}^{2}}}+|\eta|^{2}\varepsilon \bigg )^{-1} .&
\end{align}
}\normalsize
Maximizing (\ref{obj_fun2}) corresponds to maximizing $\zeta(\mathbf{W}_{r},\mathbf{W}_{c})$, or equivalently, $\mathbf{a}^{H}\mathbf{R}_{\mathrm{x}}\mathbf{a}$. Consequently, problem (\ref{p1}) is reformulated into problem (\ref{p2}), thereby concluding the proof.

\end{appendices}

\end{document}